\documentclass[preprint,12pt,authoryear]{elsarticle}

\usepackage{amssymb}
\usepackage{amsmath}
\usepackage{amsmath}
\usepackage{epsfig}
\usepackage{color}

\usepackage{mathrsfs}
\usepackage{stmaryrd}
\usepackage{setspace}
\usepackage{color}
\usepackage{bm}
\usepackage{subcaption}
\usepackage{graphicx}
\usepackage{multirow}
\usepackage{makecell}
\usepackage{amsmath}
\usepackage{caption}
\usepackage{subcaption}
\usepackage{amssymb}

\usepackage{hyperref}
\usepackage{amsmath}
\usepackage{bm}

\usepackage{booktabs}

\usepackage{tikz-cd}

	\usepackage{amsmath}
	\usepackage{graphicx}
	\usepackage{enumerate}
	\usepackage{natbib} 
	\usepackage{url} 
	\usepackage{bm}
\usepackage{subcaption}
\usepackage{graphicx}
\usepackage{multirow}
\usepackage{algorithm}
\usepackage{algorithmic}
\usepackage{amsthm}

\newtheorem{theorem}{Theorem}

\theoremstyle{definition}
\newtheorem{remark}{Remark}
\theoremstyle{definition}

\theoremstyle{definition}
\newtheorem{assumption}{Assumption}

\usepackage{times}
\usepackage{bm}
\usepackage{natbib}

\makeatletter

\def\@algocf@capt@plain{top}

\journal{JSPI}

\begin{document}

\begin{frontmatter}



\title{Fixed-budget optimal designs for multi-fidelity computer experiments}


\author{Gecheng Chen and Rui Tuo}

\affiliation{organization={Wm Michael Barnes ’64 Department of Industrial \& Systems Engineering},
            addressline={Texas A\&M University}, 
            city={College Station},
            postcode={77845}, 
            state={Texas},
            country={U.S.}}

\begin{abstract}
This work focuses on the design of experiments of multi-fidelity computer experiments. We consider the autoregressive Gaussian process model proposed by Kennedy and O'Hagan (2000) and the optimal nested design that maximizes the prediction accuracy subject to a budget constraint. An approximate solution is identified through the idea of multi-level approximation and recent error bounds of Gaussian process regression. The proposed (approximately) optimal designs admit a simple analytical form. We prove that, to achieve the same prediction accuracy, the proposed optimal multi-fidelity design requires much lower computational cost than any single-fidelity design in the asymptotic sense. Numerical studies confirm this theoretical assertion.
\end{abstract}

\begin{highlights}
\item This work introduces a modified autoregressive model to formulate computer code with infinite fidelity levels.
\item This work proposes a fixed precision optimal design for the modified autoregressive models.
\item This work analyzes the upper bound of the total cost for the proposed MLGP design and compares it with the single fidelity designs.
\end{highlights}


\end{frontmatter}


\section{Introduction}


With the advance of mathematical modeling and numerical computation, computer simulations have been widely used to mimic complex physical systems. Compared with physical experiments, computer models are often much less costly and more flexible to run. 
Conceptually, each computer model is based on a mathematical model in terms of, for instance, a set of partial differential equations. The aim of a computer code run is to find a numerical solution to this mathematical model. The design and analysis of computer experiments, an area drawing increasing attention recently, aim to use statistical approaches to collect and analyze computer simulation responses and facilitate decision-making. We refer to the books by \cite{santner2003design,gramacy2020surrogates} for an introduction to this field. One of the most important tasks in this area is to build \textit{surrogate models} of computer simulations. A surrogate model is a statistical model constructed using the computer outputs over a selected set of input sites. A point evaluation of a surrogate model is much less costly than running the corresponding computer simulation code. Thus we can use the surrogate model to explore and analyze the underlying computer response function, which expedites the decision-making.

Many computer simulation codes allow users to specify an accuracy level of the numerical responses, in terms of the degree of discretization, or the number of iterations, etc. Generally, there is a trade-off between the accuracy and the computational cost: a high-accuracy computer run is more costly than a low-accuracy run \citep{qian2008bayesian,xiong2013sequential}. 
Many statisticians have considered the problem of integrating the computer responses from multiple accuracy (also known as fidelity) levels to make predictions. The first model was suggested by \cite{kennedy2000predicting}, who proposed a Gaussian-process-based autoregressive model to integrate the output data from computer experiments with multiple fidelity levels. This model has been widely used in real-world applications, e.g., simulation oil pressure \citep{kennedy2000predicting},  designing of linear
cellular alloy \citep{xiong2013sequential} and prediction of band gaps \citep{pilania2017multi}. Alternative models multi-fidelity computer experiments include Bayesian hierarchical models \citep{qian2008bayesian}, deep Gaussian process model \cite{cutajar2019deep}, and nonstationary Gaussian process models \citep{tuo2014surrogate}. In this article, we focus on the autoregressive models given the wide acceptance of these models.

Despite the widespread use of multi-fidelity computer experiments, there remain some unanswered questions. First, it is unclear why one would prefer a multi-fidelity computer experiment over a simple single-fidelity experiment. A conventional explanation says that the low-fidelity data can help explore the underlying response function quickly, and the high-fidelity data are then used to refine the model. 
This assertion is usually confirmed by numerical studies, but there is a lack of quantitative analysis of this relationship.

The second question is regarding the choice of the set of input sites for the computer simulation, known as a design of experiment. Studying the design of computer experiments is considered important because each computer run can be costly, and a good design can help collect more information under the same computational cost \citep{santner2003design,fang2005design}. The importance of design of experiments is even more dramatic for multi-fidelity computer experiments, because a poor design can even have an adverse effect on the performance. 


A number of designs for multi-fidelity computer experiments are proposed in the literature. 
\cite{qian2009nested} proposed the nested Latin hypercube designs (NLHDs). The design points of an NLHD on one fidelity level form a Latin hypercube, and those from different fidelity levels are nested.
\cite{he2011nested} proposed the nested orthogonal array-based Latin hypercube designs to achieve stratification in both bivariate and univariate margins. \cite{rennen2010nested} constructed nested maximin Latin hypercube designs using a numerical search algorithm. See also \cite{qian2009nested,qian2012sliced,hwang2016sliced,qian2006building,tuo2014surrogate,xiong2013sequential,he2011nested,ehara2021adaptive,sung2022stacking} for more related works.  
Despite the rich literature, none of these works addressed the aforementioned questions. Some existing works provided theoretical justifications in terms of numerical integration, e.g., \cite{he2016central}, but it is still not clear why these designs are suitable for the autoregressive models.

There is another class of approaches from the area of applied mathematics for a related problem, known as multi-level Monte Carlo (MLMC) \citep{giles2008multilevel,giles2015multilevel,heinrich2001multilevel}. These methods 
are algorithms for computing expectations that arise in stochastic simulations with different levels of accuracy.
MLMC reduces the computational cost of the traditional Monte Carlo method by taking most samples with low accuracy, and only a few samples taken at high accuracy. Multifidelity Monte Carlo (MFMC) \citep{ng2014multifidelity}, based on the control variable techniques for Monte Carlo simulation, resembles the MLMC method for more general engineering models. More related works can be found in \cite{haji2016multi,peherstorfer2016multifidelity,peherstorfer2016optimal,geraci2017multifidelity,teckentrup2015multilevel}. Even though these multi-level based designs are equipped with rigorous mathematical foundations, they are not based on Gaussian process models so that they cannot be used for posterior-based analysis and uncertainty quantification. 



This work aims at underpinning the multi-fidelity computer experiments by analyzing their predictive error theoretically. Inspired by the idea of multi-level methods, we formulate a model framework for multi-fidelity computer experiments. We then consider the fixed precision optimal design defined as the design with the minimum total simulation cost subject to an accuracy guarantee. Our main theoretical contribution is to obtain the order of magnitude of the simulation cost as the required predictive error of the autoregressive model goes to zero. The analysis is based on the error bounds for Gaussian process regression provided by \cite{tuo2020kriging}. We prove that a well-designed multi-fidelity experiment will require much lower simulation cost in the order of magnitude to reach the same accuracy level, compared with the best single-fidelity experiment. This result confirms the superiority of multi-fidelity computer experiments. Based on the theory, we propose a novel design scheme referred to as the multi-level Gaussian process (MLGP) designs. Compared with the existing optimal design methods, the proposed method is easy to implement because it does not require a numerical search. Numerical studies confirm the superiority of the proposed method over existing alternatives.

The remaining part of this work is organized as follows. Section \ref{introtoauto} reviews the autoregressive models. In Section \ref{infinite}, we introduce the formulation of the modified autoregressive model and in Section \ref{theory}, we introduce the formulation of the modified autoregressive model and propose the MLGP design method. In Section \ref{algorithm}, we give the detailed steps for implementing the MLGP method in applications. In Section \ref{numerical_study}, we conduct some numerical studies, which show the nice performance of the proposed MLGP. The conclusions and some discussion are provided in Section \ref{conclusion}. Technical proofs are presented in the Appendix.

\section{Review on Gaussian process regression and autoregressive models}\label{introtoauto}
In this section, we review the Gaussian process regression and the autoregressive models for multi-fidelity computer experiments.

\subsection{ Gaussian process regression}

Let $Z$ be a Gaussian process on $\mathbb{R}^d$ with mean zero and covariance function $r(x_1, x_2)$. We call $Z$ stationary \citep{santner2003design} if $r$ can be expressed as a function of $x_1 - x_2$, i.e.,
$r(x_1, x_2) = \sigma^2\Phi(x_1 - x_2),$
where $\sigma^2$ is the variance, and $\Phi$ is a correlation function with $\Phi(0)=1$; otherwise, we call it nonstationary. Given scattered evaluations $(x_1,Z(x_1)),\ldots,(x_n,Z(x_n))$, the Gaussian process regression method (also known as kriging) estimates $Z(x)$ for any $x\in\mathbb{R}^d$ using the conditional expectation
    \begin{eqnarray}\label{gp}
    \hat{Z}(x):=E(Z(x)|Z(x_1),\ldots,Z(x_n))=a^T(x)R^{-1}Y,\label{GPR}
    \end{eqnarray}
where $a(x):=(r(x-x_1),\ldots,r(x-x_n))^T, R=(r(x_j-x_k))_{j k}$ and $Y=(Z(x_1),\ldots,Z(x_n))^T$. In the area of computer experiments, (\ref{gp}) is used to reconstruct the response function $Z(x)$, where $\{x_1,\ldots,x_n\}$ denotes the set of designs points, $\{Z(x_1),\ldots,Z(x_n)\}$ the corresponding response, and $x$ an untried point.

\subsection{Autoregressive model}

\cite{kennedy2000predicting} proposed a model to fit multi-fidelity computer experiments using Gaussian process regression. The main objective of this model is to incorporate the computer outputs with different fidelity levels and predict the response value of the computer code at an untried point of the highest fidelity level.

Suppose we have $K$ levels of computer codes, with increasing accuracies, denoted as $y_1, \dots, y_K$. \cite{kennedy2000predicting} considered the following autoregressive model to link each pair of consecutive $y_i$'s:
\begin{eqnarray}\label{auto}
    y_i(x) = \rho_{i-1}y_{i-1}(x) + \delta_i(x), \quad for \quad i=2,\dots, K,
\end{eqnarray}
where $\rho_{i}$'s are known or unknown autoregressive coefficients, and $y_1$ and $\delta_t$'s as mutually independent Gaussian processes for all t.

This autoregressive model contains some hyper-parameters. The parameter estimation is usually done using maximum likelihood \citep{forrester2007multi} or Bayesian methods \citep{kennedy2000predicting}. Both approaches involve certain iterative algorithms and thus are expensive in computation. Here we skip the details of the hyper-parameters estimation because they are irrelevant to our main topic. We only consider the prediction for the computer outputs at the highest fidelity given the hyper-parameters.  The autoregressive model has been widely used in real-world applications for multi-fidelity computer experiments,  e.g., simulation oil pressure \citep{kennedy2000predicting},  designing of linear
cellular alloy \citep{xiong2013sequential} and prediction of bandgaps \citep{pilania2017multi}.

\section{Autoregressive models for computer code with infinite fidelity levels}\label{infinite}

In this work, we focus on computer experiments with infinite fidelity levels. There are many examples of this kind. For example, in finite element analysis, the accuracy is governed by the mesh density, which, at least conceptually, has infinite levels. Other examples include the step length in a finite difference algorithm, the number of iterations in an iterative algorithm, etc. Such a parameter is referred to as a ``tuning parameter'' in \cite{tuo2014surrogate}.

\begin{figure}[ht]
    \centering
    \includegraphics[width=\textwidth]{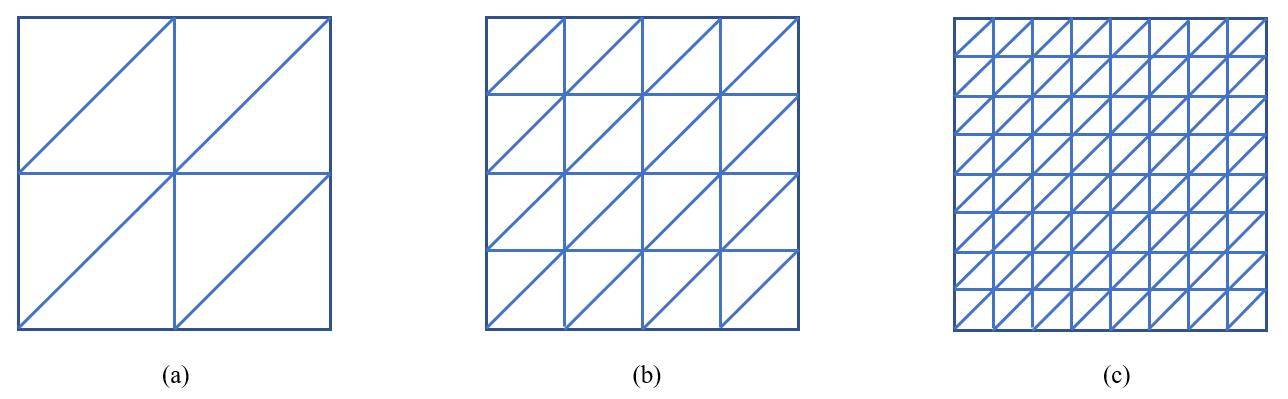}
    \caption{Consecutively refined meshes. The number of the triangle local patches quadruples as the mesh size is halved.}
    \label{fig:mesh}
\end{figure}

Let us use the finite element method (FEM) as an example to introduce our intuition. 
FEM generally uses piecewise linear functions over prespecified local patches to approximate general functions, such as the partial differential equation (PDE) solutions. The number of local patches is governed by the chosen mesh size. A typical multi-fidelity computer experiment uses a sequence of consecutively refined meshes. Figure \ref{fig:mesh} shows the finite elements corresponding to the first three mesh sizes in the set of refining meshes $\{1/2, 1/4, 1/8,...,1/2^n,...\}$ in a unit square.

Here comes the question: \textit{what is a suitable configuration of autoregressive models for the above FEM example which can potentially have infinite fidelity levels?} 
Clearly, we need to make some modifications to (\ref{auto}).
The most obvious one follows from the fact that there are infinite fidelity levels in the above example. That is, conceptually, we should have computer response functions $y_i$ for $i=1,\ldots,K,\ldots$. Next, we presume that the FEM solution converges to the exact solution of the PDE as the mesh size goes to zero, which is usually the case. Then the underlying quantity of interest is the function $y_\infty:=\lim_{i\rightarrow\infty}y_i$.
To guarantee the existence of this limit, more constraints must be imposed in the autoregressive model (\ref{auto}). First, we must have $\lim_{i\rightarrow \infty}\rho_i=1$ as the existence of $y_\infty$ requires $\lim_{i\rightarrow \infty}(y_i-y_{i-1})=0$. In this work, we assume $\rho_i=1$ for all $i$ for simplicity. Then we have the expression
$$y_\infty=y_1+\sum_{i=2}^\infty \delta_i. $$

Note that $\text{Var}(y_\infty)=\text{Var}(y_1)+\sum_{i=2}^\infty \text{Var}(\delta_i)$. To ensure that $y_\infty$ has a finite variance, $\delta_i$ must have decreasing variances. Here we model $\delta_i \sim GP(0, \lambda^{2i}\sigma^2\Phi_i)$ for some correlation function $\Phi_i$, variance $\sigma^2$ and $0<\lambda<1$. The exponential decay in the variance can be justified because the mesh size in Figure \ref{fig:mesh} also decays exponentially as the refinement continues. Since we have not seen any data in the experimental design stage, for simplicity, we assume that all the $\delta_i$'s share the same correlation function $\Phi$.

We now define the modified version of the autoregressive model as, 
\begin{eqnarray}\label{newauto}
    y_i(x) = y_{i-1}(x) + \delta_i(x), \quad for \quad i=0,\dots, K, \dots,
\end{eqnarray}
where $y_{-1} = 0$, and $\delta_i \sim GP(0, \lambda^{2i}\sigma^2\Phi)$ are mutually independent.

\section{Multiple fidelity designs based on MLGP}\label{theory}

It is hopeful that a multi-fidelity experiment can perform better than a single-fidelity experiment, in the sense that the computational cost is lower to achieve a fixed predictive accuracy. Otherwise, there is no value in using a multi-fidelity design. In this section, the goal is to understand the best prediction performance of a multi-fidelity experimental design.

First, we introduce an assumption for the computational cost of each experimental trial.
At the same time, the number of finite elements increases exponentially as the mesh refines iteratively so that the resulting cost for each untried point also increases exponentially. In view of this, we assume the cost for each sample in the $i$-th level as $C_b a^i \leq C_i \leq C_a a^i,$ where $C_a$ and $C_b$ are constants, $a>1$ is the cost ratio. Additionally, the difference in the approximation values given by successive mesh sizes decreases as the mesh refines.


\subsection{A fixed precision optimal design for multi-fidelity models}

In this section, we introduce the problem setting for constructing optimal designs for the modified autoregressive model with fixed precision. As discussed in the above section, we use $y_\infty$ to denote the underlying function. (\ref{newauto}) allows us to reduce the multi-fidelity problem to a sequence of single-fidelity problems. If we can reconstruct $\delta_i$ (i.e., $y_i - y_{i-1}$) for each $i$, we are able to reconstruct $y_\infty$ using the summation of all these constructed functions. To reconstruct $\delta_i$, a simple idea is to evaluate the function $\delta_i$ over a set of selected points, denoted as $\mathcal{X}_i$ correspondingly, via the Gaussian process regression. We assume the underlying function in the $K$-th layer, i.e., $ y_K = \sum_{i=0}^{K} \delta_i$, is sufficiently accurate. One can reconstruct the $y_K$ by $\hat{y}_K$ defined as (\ref{originalapp}) to approximate $y_\infty$.
\begin{eqnarray}\label{originalapp}
    \hat{y}_K(x) = E\left[ \sum_{i=0}^{K}\delta_i(x) \Big| y_K(\mathcal{X}_K),...,y_0(\mathcal{X}_0) \right]
\end{eqnarray}

Denote the desired accuracy level, in terms of the lack of fit, as $\epsilon$ ($\epsilon<1$). Then we can formulate the general objective function for constructing optimal designs for the modified autoregressive model as 
\begin{alignat}{2}
\min_{n_i, K}  \quad & \sum_{i=0}^K n_iC_i \label{objectve}\\
\mbox{s.t.}\quad
&E\left\|y_\infty - \hat{y}_K\right\|^2_{L^2} \leq \epsilon^2, \label{constraint}
\end{alignat}
where $n_i\in \mathbb{N} $ denotes the sample size of $\mathcal{X}_i$, and $\left\|y_\infty - \hat{y}_K\right\|^2_{L^2}$ is the expected squared predictive error of $\hat{y}_K$. This objective function can be interpreted as: we want to minimize the cost caused by the designs $\{n_i, K\}$ which satisfies the accuracy requirement. 

\subsection{MLGP method}


The main difficulties for solving (\ref{objectve}) include: 1. $y_\infty$ is unknown; 2. $\hat{y}_K$ shown in (\ref{originalapp}) does not have a closed form under an arbitrary design; 3. the integer optimization problem is NP-hard. In this section, we make some approximations and restrictions to make (\ref{objectve}) solvable. In detail, to make the expected squared error in (\ref{constraint}) tractable, we only consider nested designs \cite{kennedy2001bayesian,qian2009nested} to simplify $\hat{y}_K$. Additionally, we replace the left-hand side of (\ref{constraint}) with its upper bound to get an approximated solution. We also relax $n_i$'s and $K$ to be real numbers at first to search for the optimal solution, and then round down them to the final optimal solution.

The widely used nested designs can refine the low-accuracy experiments with the high-accuracy experiments to obtain a better model because the response is also available in low-accuracy experiments at the same design points \citep{qian2009nested}. Denote $\mathcal{X}_0, \mathcal{X}_1, \ldots, \mathcal{X}_K$ as a sequence of nested sets of points, then the correlation between them can be formulated as
$\mathcal{X}_K \subset \cdots \subset \mathcal{X}_1 \subset \mathcal{X}_0$. A good property of the nested design given in Theorem \ref{theorem1} helps simplify the approximation (\ref{originalapp}). 

\begin{theorem}
\label{theorem1}
Suppose mutually independent $\delta_i \sim GP(0, \sigma^2\Phi_i )$, $i=0,1,\dots,K$, where $\sigma^2$ is the variance, $\Phi$ is the correlation function and $0<\lambda^2<1$. Let $\mathcal{X}_K \subset \cdots \subset \mathcal{X}_1 \subset \mathcal{X}_0$ be a sequence of nested sets of points. Then for each $x$, the estimator $\hat{y}_K(x)$ defined in (\ref{originalapp}) admits the expression
\begin{eqnarray}\label{approximationnew}
    \hat{y}_K(x) =  \sum_{i=0}^{K}E\left[(y_i-y_{i-1})(x)\Big|(y_i-y_{i-1})(\mathcal{X}_i)\right].
\end{eqnarray}

\end{theorem}

Theorem \ref{theorem1} shows that if a nested design is used, Gaussian process regression for an autoregressive model can be done by solving
a set of single-fidelity Gaussian process regression problems. Specifically, (\ref{approximationnew}) is a sum of $K+1$ GP predictors.
Theorem \ref{theorem1} gives us a good incentive to use nested designs, and then the multi-level interpolation approximation (\ref{originalapp}) can be rewritten as (\ref{approximationnew}).

The next step is to identify the optimal nested designs for the (\ref{objectve}). 
Because $\hat{y}_K$ depends only on $y_0,\ldots,y_K$, we can see that $y_\infty-y_K$ is independent of $y_K-\hat{y}_K$ and therefore,   
\begin{eqnarray}\label{error}
    E\left\|y_\infty - \hat{y}_K\right\|^2_{L^2} = E\left\|y_\infty - y_K\right\|^2_{L^2} + E\left\|y_K - \hat{y}_K\right\|^2_{L^2}.
\end{eqnarray}

To achieve the desired accuracy, a simple idea is to bound each of the right-hand side terms of (\ref{error}) by $\epsilon^2/2$, i.e., we convert the constrain (\ref{constraint}) into
\begin{eqnarray}\label{newconstrain1}
    E\left\|y_\infty - y_K\right\|^2_{L^2} <\epsilon^2/2,
\end{eqnarray}
and
\begin{eqnarray}\label{newconstrain2}
    E\left\|y_K - \hat{y}_K\right\|^2_{L^2} < \epsilon^2/2.
\end{eqnarray}
where (\ref{newconstrain1}) reflects the error caused by using $y_K$ to approximate $y_\infty$ and (\ref{newconstrain2}) reflects the error caused by the Gaussian process interpolation. Next, we will handle these two parts separately.

First, we focus on (\ref{newconstrain1}). Given the modified autoregressive model with shrinking variance, it is easy to compute the upper bound of the error of the left-hand side of (\ref{newconstrain1}).
\begin{theorem}
\label{theorem2}
Given mutually independent $\delta_i \sim GP(0, \lambda^{2i}\sigma^2\Phi)$, $i=0,1,\dots,K,\dots$, with $0<\lambda^2<1$ and $\sigma^2>0$, we have 
\begin{eqnarray}\label{thm2}
    E\left\|y_\infty - y_K\right\|_{L^2}^2 \leq C_l^2\lambda^{2K}\sigma^2,
\end{eqnarray}
where $C_l = \sqrt{C_s}/(1-\lambda)$ is a constant, and $C_s$ is the volume of the region $\mathcal{X}$. 
\end{theorem}

Based on Theorem \ref{theorem2}, it is sufficient to bound the upper bound of the left-hand side of (\ref{thm2}) by $\epsilon^2/2$ to make (\ref{newconstrain1}) hold, that is,
\begin{align*}
    E\left\|y_\infty - y_K\right\|^2_{L^2} \leq C_l^2\lambda^{2K}\sigma^2 \leq \epsilon^2 /2.
\end{align*}

Then we can choose 
\begin{eqnarray}\label{k}
    K = \left\lceil \log_{\lambda^2} \frac{\epsilon^2}{2C_l^2\sigma^2} \right\rceil,
\end{eqnarray}
as the total number of levels $K$, where 
$\left\lceil a \right\rceil$ denotes the largest integer which is smaller or equal to $a$.

Next, we turn to (\ref{newconstrain2}), which is associated with the error from the Gaussian process modeling. Here we make a similar assumption as the \cite{tuo2020kriging} to our nested designs. Before introducing the assumption, we make some definitions. For a set of design points $\mathcal{X} = \{x_1, x_2,...,x_n\}\subset \Omega$, define the fill distance as
$$h_{\mathcal{X}, \Omega}=\sup_{x\in \Omega} \inf_{x_j \in \mathcal{X}}||x-x_j||,$$
and the separation radius as
$$q_\mathcal{X} = \min_{1\leq j\neq k\leq n}||x_j-x_k||/2.$$
\begin{assumption}\label{assumption1}
$\mathcal{X}_K \subset ... \subset \mathcal{X}_1 \subset \mathcal{X}_0$ is a sequence of nested sets of points, denote the corresponding fill distance and separation radius as $h_{\mathcal{X}_i, \Omega}$ and $q_{\mathcal{X}_i}$. There exists a constant $A$ such that
$h_{\mathcal{X}_i, \Omega} / q_{\mathcal{X}_i} \leq A$ holds for all $i$.
\end{assumption}
Given  scattered evaluations $(x_1,Z(x_1)),\ldots,(x_n,Z(x_n))$ where the design is quasi-uniform, if we use the Mat\'ern kernel to reconstruct the underlying function $Z(x)$, then there exist two positive constants $p$ and $q$ such that \cite{tuo2020kriging}:
\begin{equation}
E\left\| Z(x) - \hat{Z}(x)\right\|^2_{L^2} <  pn^{-2\nu/d},
\end{equation}
and
\begin{eqnarray}
E\left\| Z(x) - \hat{Z}(x)\right\|^2_{L^2} >  qn^{-2\nu/d},
\end{eqnarray}
where $\hat{Z}(x)$ is the reconstructed function, $\nu$ is the smoothness parameter of the Mat\'ern kernel and $d$ is the dimension of the design. 

Based on the error bound of the original Gaussian process regression, we propose the upper bound of the interpolation error in the modified autoregressive model in Theorem \ref{theorem3}.  
\begin{theorem}
\label{theorem3}
Assume a infinite series $\delta_i \sim GP(0, \lambda^{2i}\sigma^2\Phi)$ and they are mutually independent. Denote $y_K$ as the summation of the first $(K+1)$ terms, and $\hat{y}_K$ defined in (\ref{approximationnew}) as the emulation of $y_K$ based on the set of points $\mathcal{X}_i$ and Mat\'ern kernel, then under Assumption \ref{assumption1} we have
\begin{eqnarray}\label{thm3}
    E\left\|y_K - \hat{y}_K\right\|^2_{L^2} \leq \sum_{i=0}^K p\lambda^{2i}\sigma^2n_i^{-2\nu/d},
\end{eqnarray}
where $n_i$ is the number of points of $X_i$, $d$ is the number of dimension, $p$ is a constant, $\nu$ is the smoothness parameter of $\Phi$.

\end{theorem}

Based on Theorem \ref{theorem3}, it is sufficient to bound the upper bound of the left-hand side of (\ref{thm3}) by $\epsilon^2/2$ to make (\ref{newconstrain2}) hold. 
Then we can approximate the solution of the objective function (\ref{objectve}). Based on \cite{tuo2020kriging}, we state the upper bound of the interpolation error in the modified autoregressive model in Theorem \ref{theorem3}.  
\begin{theorem}
\label{theorem3}
Assume a infinite series $\delta_i \sim GP(0, \lambda^{2i}\sigma^2\Phi)$ and they are mutually independent. Denote $y_K$ as the summation of the first $(K+1)$ terms, and $\hat{y}_K$ defined in (\ref{approximationnew}) as the emulation of $y_K$ based on the set of points $\mathcal{X}_i$ and Mat\'ern kernel, then under Assumption \ref{assumption1} we have
\begin{eqnarray}\label{thm3}
    E\left\|y_K - \hat{y}_K\right\|^2_{L^2} \leq \sum_{i=0}^K p\lambda^{2i}\sigma^2n_i^{-2\nu/d},
\end{eqnarray}
where $n_i$ is the number of points of $X_i$, $d$ is the number of dimension, $p$ is a constant, $\nu$ is the smoothness parameter of $\Phi$.
\end{theorem}

Based on Theorem \ref{theorem3}, it is sufficient to bound the upper bound of the left-hand side of (\ref{thm3}) by $\epsilon^2/2$ to make (\ref{newconstrain2}) hold. 
Then we can approximate the solution of objective function (\ref{objectve}) with (\ref{constraint}) by solving a relaxed version, i.e., 
\begin{equation}\label{newnewobjectve}
    \begin{split}
        \min_{n_i} \quad &\sum_{i=0}^K n_iC_i \\
\mbox{s.t.}\quad
&\sum_{i=0}^K p\lambda^{2i}\sigma^2n_i^{-2\nu/d} \leq \epsilon^2/2
    \end{split}
\end{equation}
where $K = \lceil 2\log_{\lambda^2} \frac{\epsilon}{2C_l\sigma} \rceil$. One can see that (\ref{newnewobjectve}) is an integer programming problem which is NP-hard. In view of this, we consider a relaxed version of it. First we assume $n_i$'s can choose any real number, then (\ref{newnewobjectve}) becomes a convex optimization problem whose optimal solution is
\begin{eqnarray}\label{originalsolution}
    n_i =  \left(\frac{\epsilon^2}{2p\sigma^2S}\right)^{-\frac{d}{2\nu}}\left(\frac{a^i}{\lambda^{2i}}\right)^{-\frac{d}{d+2\nu}} ,\quad i=0,...,K, 
\end{eqnarray}
where $S = \sum_{i=0}^K(a^i)^{\frac{\nu}{d+2\nu}}(\lambda^{2i})^{\frac{d}{d+2\nu}}$, $p$ is a constant. The proof of (\ref{originalsolution}) is given in the Appendix.

Note that $\left\lceil n_i \right\rceil$'s also locate in the feasible region of (\ref{newnewobjectve}). We regard the $\left\lceil n_i \right\rceil$'s shown in (\ref{ni}) as a nearly optimal solution to (\ref{newnewobjectve}).
\begin{eqnarray}\label{ni}
    n_i = \left\lceil \left(\frac{\epsilon^2}{2p\sigma^2S}\right)^{-\frac{d}{2\nu}}\left(\frac{a^i}{\lambda^{2i}}\right)^{-\frac{d}{d+2\nu}} \right\rceil,\quad i=0,...,K, 
\end{eqnarray}

In summary, we consider the designs generated based on (\ref{ni}) and (\ref{k}) as the optimal designs for the modified autoregressive model. We will discuss the implementation details in Section \ref{algorithm}.

\subsection{Comparison with single-fidelity experiments}
In this section, we will show that the MLGP needs less cost than any single-level method to achieve a given accuracy provided that the accuracy level is small enough. Let us denote $a = \exp{(\alpha)}$ and $\lambda^2 = \exp{(-\beta)}$ for convenience ($\alpha, \beta>0$). In this section, we use the symbol ``$\lesssim$" to represent that the left-hand side is less than the right-hand side multiplied by a constant depending only on $\Phi$. First, we give an upper bound for the total cost for the MLGP method that reaches a fixed accuracy level $\epsilon$.

\begin{theorem}\label{thm4}
Assume $ y_i(x) = y_{i-1}(x) + \delta_i(x)$, for $i=0,\dots, K, \dots$, where $\delta_i {\sim} GP(0, \lambda^{2i}\sigma^2\Phi )$ and they are mutually independent, $\lambda^2<1$ and $y_{-1} = 0$. If we use the design (\ref{k}) and (\ref{ni}) to reach a fixed accuracy level $\epsilon$, then the upper bound for the total cost is summarized as
\begin{eqnarray}\label{MLGPcost}
    C_\epsilon\lesssim\left\{\begin{matrix}
 \epsilon^{-d/\nu},&2\alpha\nu < d\beta \\ 
 \epsilon^{-d/\nu}|\ln \epsilon^2|^{\frac{d+2\nu}{2\nu}}, & 2\alpha\nu = d\beta\\ 
 \epsilon^{-2\alpha/\beta},&2\alpha\nu > d\beta 
\end{matrix}\right.
\end{eqnarray}

\end{theorem}

Next, we discuss the lower bound for the total cost if we implement the single-fidelity method to reach the accuracy level $\epsilon$. 
\begin{theorem}\label{thm5}
Assume $ y_i(x) = y_{i-1}(x) + \delta_i(x)$, for $i=0,\dots, K, \dots$, where $\delta_i {\sim} GP(0, \lambda^{2i}\sigma^2\Phi )$ and they are mutually independent, $\lambda^2<1$ and $y_{-1} = 0$. If we only spread design samples in one layer to reach a fixed accuracy level $\epsilon$, then the lower bound for the total cost is
\begin{eqnarray}\label{SLcost}
    C_\epsilon^{SL} \gtrsim \epsilon^{-(\frac{d}{\nu} + \frac{2\alpha}{\beta}) }.
\end{eqnarray} 
\end{theorem}

\section{Fixed budget designs}\label{algorithm}
For many real applications, a more realistic assumption is that the total budget, rather than the accuracy level, is fixed. 
In this section, we will introduce how to implement the MLGP design method given a fix budget.

First, assume we know the values for the hyper-parameters $\lambda^2$, correlation function, and $\nu$. In practice, $\lambda^2$ can be decided based on expert knowledge. For example, we can choose the decreasing ratio of the mesh sizes as $\lambda^2$ in the FEM case. The correlation function and the corresponding parameter(s) need to be decided based on the smoothness of the underlying function, see  \cite{wilson2013gaussian,schulz2018tutorial} for more information.

In order to obtain the designs based on (\ref{k}) and (\ref{ni}) under a certain budget $B$, we need to have the exact values for the parameters $p$, $C_l$, $\sigma$ and the desired accuracy level $\epsilon$. These parameters are often hard to decide in real-world applications. However, the number of available layers is often determined in real-world applications. Based on this, we just assume $K$ is determined. Please note that this assumption does not mean we need to use all the layers in our designs because $n_i$ can be very small for a certain layer if the MLGP method thinks that the corresponding layer is not needed. 

The next thing is to calculate $n_i$. Denote $\tau := p\sigma^2/\epsilon^2$ and $\eta := p/C_l^2$, then we rewrite (\ref{ni}) as
\begin{eqnarray}\label{ninew}
    n_i = \left\lceil (2\tau S)^{\frac{d}{2\nu}}(\frac{a^i}{\lambda^{2i}})^{-\frac{d}{d+2\nu}} \right\rceil,\quad i=0,1,...,K
\end{eqnarray}
where $\tau = \frac{\eta}{2\lambda^{2K}}$.

In order to avoid the estimation of those constants ($p$, $C_l$, $\sigma$ and $\epsilon$), one can tune $\eta$ to get the corresponding optimal design under the fixed budget $B$. By choosing a value for $\eta$, we can get a value for $\tau$ which leads to corresponding $n_i, i=0,...,K$ according to (\ref{ninew}). The resulting design that triggers the highest cost and meets the budget requirement can be chosen as the optimal design $n_i^*$'s. In applications, the user can try different values for $\eta$ in a pre-specified set (e.g., the integers from $1$ to $100$) to search for the optimal design $n_i^*$'s.

Note that the design $n_i^*$'s does not use all the budget in most cases because we need to choose an integer for $n_i^*$'s when implementing (\ref{ninew}). Denote the unallocated budget as
\begin{eqnarray}\label{unallocated_budget}
    B_K = B - \sum_{i=0}^K n_i^*C_i.
\end{eqnarray}

The final step is to re-allocate $B_K$ to modify the design $n^*_i, i=0,...,K$. We implement a greedy algorithm to solve the following objective function (\ref{re1}) to reallocate the budget to further reduce the error in the left-hand side of the constraint of (\ref{newnewobjectve}). 
\begin{eqnarray}\label{re1}
\min_{n_i^{**}} \quad &\sum_{i=0}^K \lambda^{2i}\sigma^2((n_i^{**})^{-2\nu/d}-(n_i^{*})^{-2\nu/d}) \label{reallocate}\\
\mbox{s.t.}\quad
&\sum_{i=0}^K (n_i^{**}-n_i^*)C_i \leq B_K 
\end{eqnarray}
Then we get the final optimal designs $n^{**}_i, i=0,...,K$.

Denote a $K$ level design as a vector $N=[n_1, ..., n_K]$, and a function $G(N^{**}) = \sum_{i=0}^K \lambda^{2i}\sigma^2((n_i^{**})^{-2\nu/d}-(n_i^{*})^{-2\nu/d})$. Also denote a vector with the $i$-th element equal to $1$ and other elements equal to $0$ (i.e., $1^i=[0, ..., 1, ...0] $) as one sample in the $i$-th layer. Now the detailed steps of the re-allocating starting from the design $N^*$ are as follows:
\begin{itemize}
    \item Search $i$ exhaustively in the range $[0,K]$ to minimize $G(N^* + 1^i)$ on the condition that $C_i \leq B_K$;
    \item Add one sample to the $i$-th layer of $N^*$, that is, $N^* = N^* + 1^i$;
    \item Calculate the unallocated budget $B_K = B_K - C_i$;
    \item Repeat the above steps until $B_K=0$ or $B_K$ cannot afford any additional sample.
\end{itemize}

\begin{remark}
  Before reaching the optimal designs, we need to determine the values for several parameters, including $a$, $C_a$, $C_b$, $B$, $K$, $\lambda$, $\nu$ and $\eta$. Some of the parameters are given or estimated by specific applications, such as $a$, $C_b$, $C_a$, $B$ and $K$. Some of the parameters need to be chosen by the users based on experience or other knowledge, i.e., $\lambda$ and $\nu$. We may misspecify them because we do not know the true values for these two parameters. However, empirical studies show that the designs generated by the proposed MLGP can still outperform existing alternatives even with misspecified values. See Section \ref{mis} for more details. As for the tuning of $\eta$, screening only over integer values should give decent tuning results. Note that we still have a re-allocation step after this search that will further modify the design, so we still have a good chance to get a nice design even if we do not reach the optimal $\eta$ exactly. 
  If need, when the optimal $\eta$ lies between two consecutive integers, we can consider a finer discretization over the search space to obtain a better $\eta$.
  Besides, some other searching methods are available to find the optimal $\eta$ more efficiently, such as the binary search algorithm which is applied in our experiments. 
\end{remark}

Given the detailed steps for calculating the $n_i^{**}$'s for each layer, we still need to decide how to spread these samples in each layer. Please recall that the error bound in Theorem \ref{theorem3} only holds when Assumption \ref{assumption1} holds. But it is very hard to maintain this assumption in applications rigorously because it involves a lot of optimization steps. we consider approximately space-filling sequences, such as Halton sequences \citep{halton1964algorithm} which is a type of low-discrepancy sequences. Halton sequences enjoy the following two advantages. First, it can give designs of any sample sizes. Second, the sequences have analytical expressions which are easy to compute. 

\subsection{Examples}\label{examples}

In this section, we visualize the designs generated by the MLGP under different budgets or different $\lambda^2$ in $[0,1]^2$. We choose $a=4$ and $C_a=C_b=1$ for the experiments in this section.

\begin{figure}[ht]
    \centering
    \includegraphics[width=0.7\textwidth]{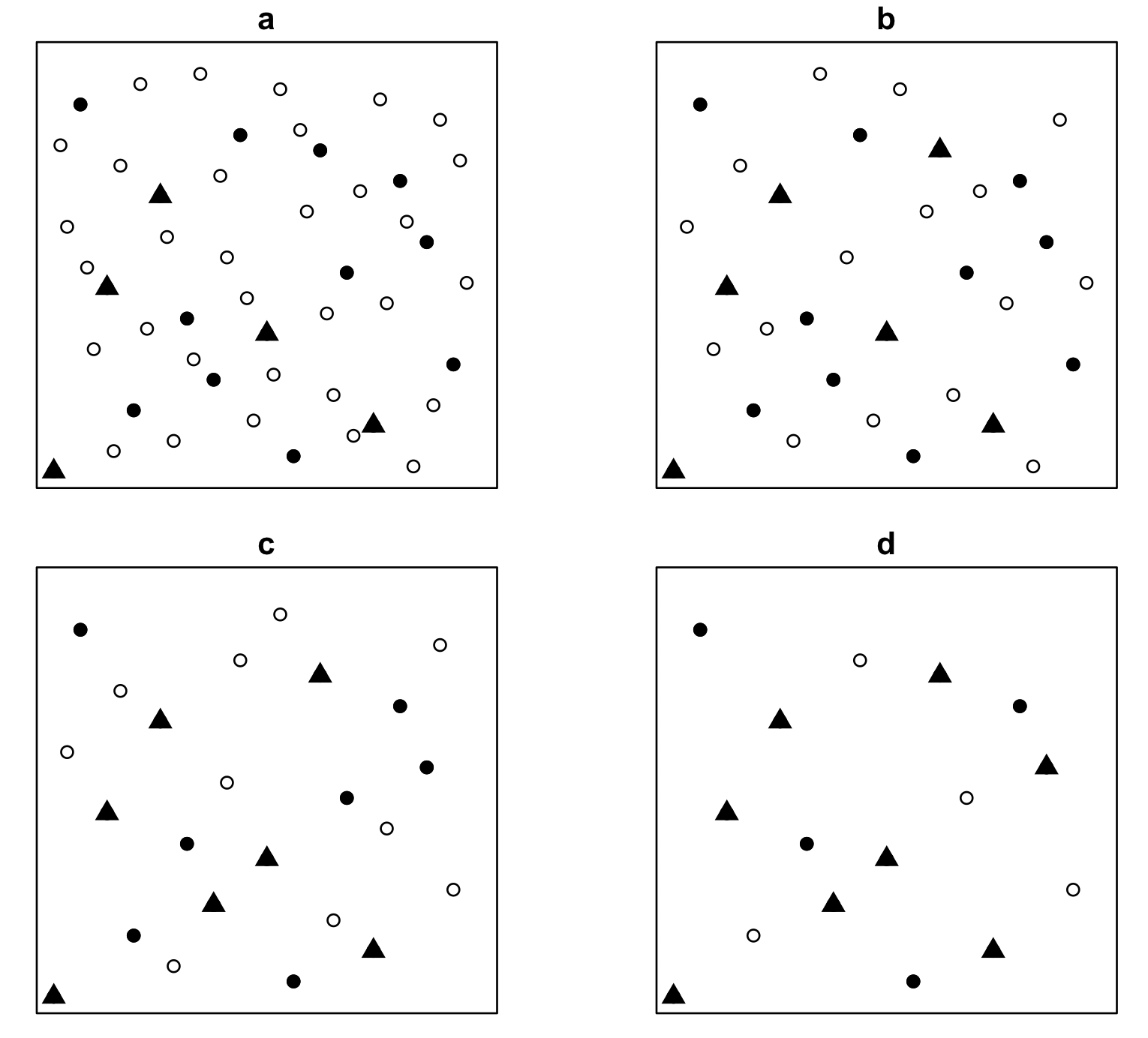}
    \caption{Designs generated by MLGP under different $\lambda^2$}
    \label{fig:2dlambda}
\end{figure}

In the first experiment, we use $B=192$, $K=2$ and Mat\'ern correlation function with $\nu = 1.25$ to generate designs under four different $\lambda^2$ using the MLGP. Tabel \ref{lambda} and Figure \ref{fig:2dlambda} show the design structures when $\lambda^2$ changes from $1/8$ to $3/4$. Note that the column "Design structure" shows the number of samples from the lowest to the highest layer. For example, the design $20,7,3$ means a nested design with $20$ samples (empty circles in Figure \ref{fig:2dlambda}) in the $0$-th layer, $7$ samples (filled circles in Figure \ref{fig:2dlambda}) in the $1$st layer and $3$ samples (filled triangles in Figure \ref{fig:2dlambda}) in the $2$nd layer. The subplot a, b, c and d are corresponding to $\lambda^2=1/8$, $\lambda^2=1/4$, $\lambda^2=1/2$ and $\lambda^2=3/4$, respectively. It can be seen that the MLGP allocates more budget on the higher layers with the $\lambda^2$ increases. This phenomenon exactly matches the changing of the true function. This is because the variance of $\delta_i$'s in (\ref{approximationnew}) in the high layers take less proportion in the underlying function $y_K$ if $\lambda^2$ is small. In other words, the information of the higher layers plays a less important role in the overall accuracy. As a result, if we want to get a low error as much as possible under a fixed budget, we should pay more budget on the samples in the lower layers and vice versa. This exclusive flexibility of the MLGP can help the users a lot to construct suitable designs when facing different $\lambda^2$ while the multi-level and single-level methods can only build designs with fixed proportions.  

\begin{figure}[ht]
    \centering
    \includegraphics[width=0.7\textwidth]{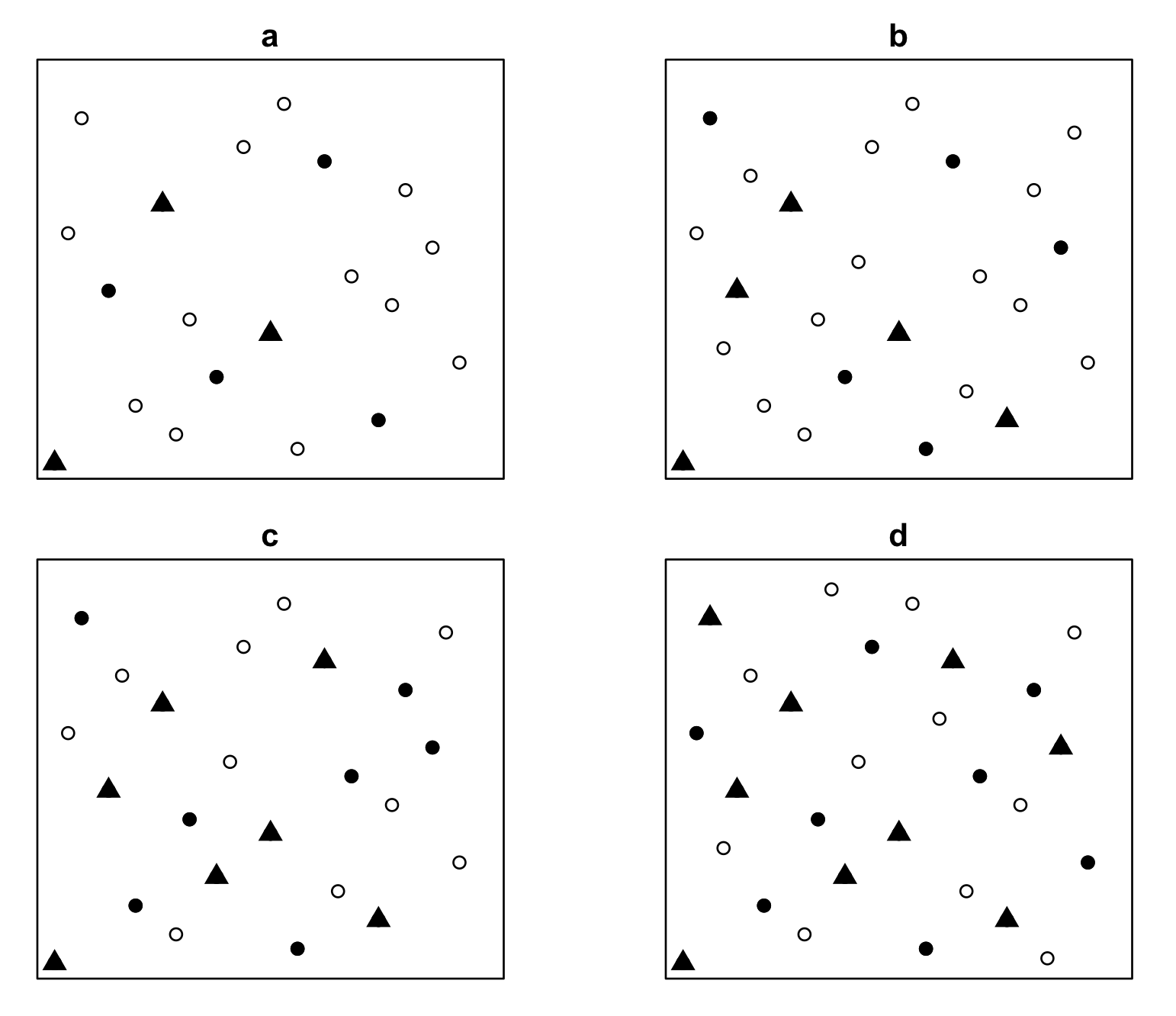}
    \caption{Designs generated by MLGP under different budgets $B$}
    \label{fig:2dcost}
\end{figure}

In the second experiment, we use $K=2$, $\lambda^2=1/2$, Mat\'ern correlation function with $\nu = 1.25$ to generate designs under budgets from $96$ to $240$ using the MLGP. Table \ref{budget} and Figure \ref{fig:2dcost} show the design structures generated by the MLGP under different $B$.

In the third experiment, we test our method under large $\nu$'s for the Mat\'ern correlation function. We use $K=3$, $\lambda = 1/2$ and budget $B=4760$ to generate designs under different $\nu$'s from $5$ to $50$. Table \ref{nu} shows the designs in this experiment, we can see that our method still works well under large $\nu$'s.

\begin{minipage}{\textwidth}
        \begin{minipage}[t]{0.5\textwidth}
            \centering
            \makeatletter\def\@captype{table}\makeatother\caption{Designs under different $\lambda^2$}
\setlength{\tabcolsep}{3mm}{
\begin{tabular}{cc}
        \toprule
            $\lambda^2$ & Design structure \\
        \midrule
            $1/8$ & 48, 16, 5 \\
        \midrule
            $1/4$ & 32, 16, 6 \\
        \midrule
            $1/2$ & 24, 14, 7 \\
        \midrule
            $3/4$ & 16, 12, 8\\
        \bottomrule
       
\end{tabular}}
\label{lambda}
        \end{minipage}
        \begin{minipage}[t]{0.5\textwidth}
        \centering
        \makeatletter\def\@captype{table}\makeatother\caption{Designs under different budgets $B$}
\setlength{\tabcolsep}{3mm}{
\begin{tabular}{cc}
 \toprule
            $B$ & Design structure \\
        \midrule
            $96$ & 20, 7, 3 \\
        \midrule
            $144$ & 24, 10, 5 \\
        \midrule
            $192$ & 24, 14, 7 \\
        \midrule
            $240$ & 28, 17, 9 \\
        \bottomrule
\end{tabular}}
\label{budget}
        \end{minipage}
                \begin{minipage}[t]{\textwidth}
        \centering
        \makeatletter\def\@captype{table}\makeatother\caption{Designs under different smoothness parameters $\nu$}
\setlength{\tabcolsep}{3mm}{
\begin{tabular}{cc}
 \toprule
            $\nu$ & Design structure \\
        \midrule
            $5$ & 24,16,16,7\\
        \midrule
            $10$ & 16, 9, 17, 7 \\
        \midrule
            $25$ & 16, 9, 9, 8 \\
        \midrule
            $50$ & 16, 9, 9, 8 \\
        \bottomrule
\end{tabular}}
\label{nu}
        \end{minipage}
    \end{minipage}

\section{Numerical study}\label{numerical_study}
In this section, some numerical studies will be conducted to show the superiority of the proposed MLGP. In Section \ref{1d}, we will first compare the performance of the proposed MLGP, classical multi-fidelity designs, and single-level designs in 1-dimensional cases when we know all hyper-parameters ($\lambda^2$, correlation function, and $\nu$). Then we will test the performance of these three methods if one of the hyper-parameters is misspecified in Section \ref{mis}. In Section \ref{2d}, we will compare the proposed MLGP with Nested Latin Hypercube designs and single-level designs in a 2-dimensional case. Emulators given by the proposed MLGP yield more accurate results than other methods in all cases.

\subsection{Numerical study in one-dimensional space}\label{1d}
In the section, we prepare a four-level function $f(x)$ (i.e., $K=3$) defined in $[0,15]$ as the simulation function. Each level is assumed as a mutually independent Gaussian process with zero mean and Mat\'ern correlation function. We assume $\lambda^2=1/3$, the correlation matrix is Mat\'ern correlation function with $\nu=1.25$. In this experiment, we choose the cost ratio $a=8$ and the cost for each sample in the $0$-th layer to be $1$, i.e., $C_a=C_b=1$.

In this experiment, we assume all the true values hyper-parameters are known. We calculate three groups of designs based on the proposed MLGP method, the classical multi-fidelity design method, and the single-level design method under different budgets. After calculating the number of samples in each layer, we generate the corresponding (nested) design samples based on the Halton sequences for all three methods. Then we train a Gaussian process regression (GPR) emulation based on each design to test the prediction accuracy on a testing set with $200$ samples. Note that we do not use maximum likelihood estimation (MLE) to estimate the variance and the length parameter. The accuracy of GPR emulations is measured by the root mean squared error (RMSE) defined as
$RMSE = \sqrt{N^{-1}\sum_{i=1}^N (y_i - \hat{y}_i)^2},$
where $N=200$, $y_i$, and $\hat{y}_i$ are the true label and the predictive label for the $i$-th testing sample, respectively.

In order to make the results more convincing and show the robustness of the proposed MLGP algorithm, we repeat this experiment $30$ times starting from different realizations of $f(x)$ and compare the mean RMSE of these methods. 

\begin{figure}[ht]
    \centering
    \includegraphics[width=0.7\textwidth]{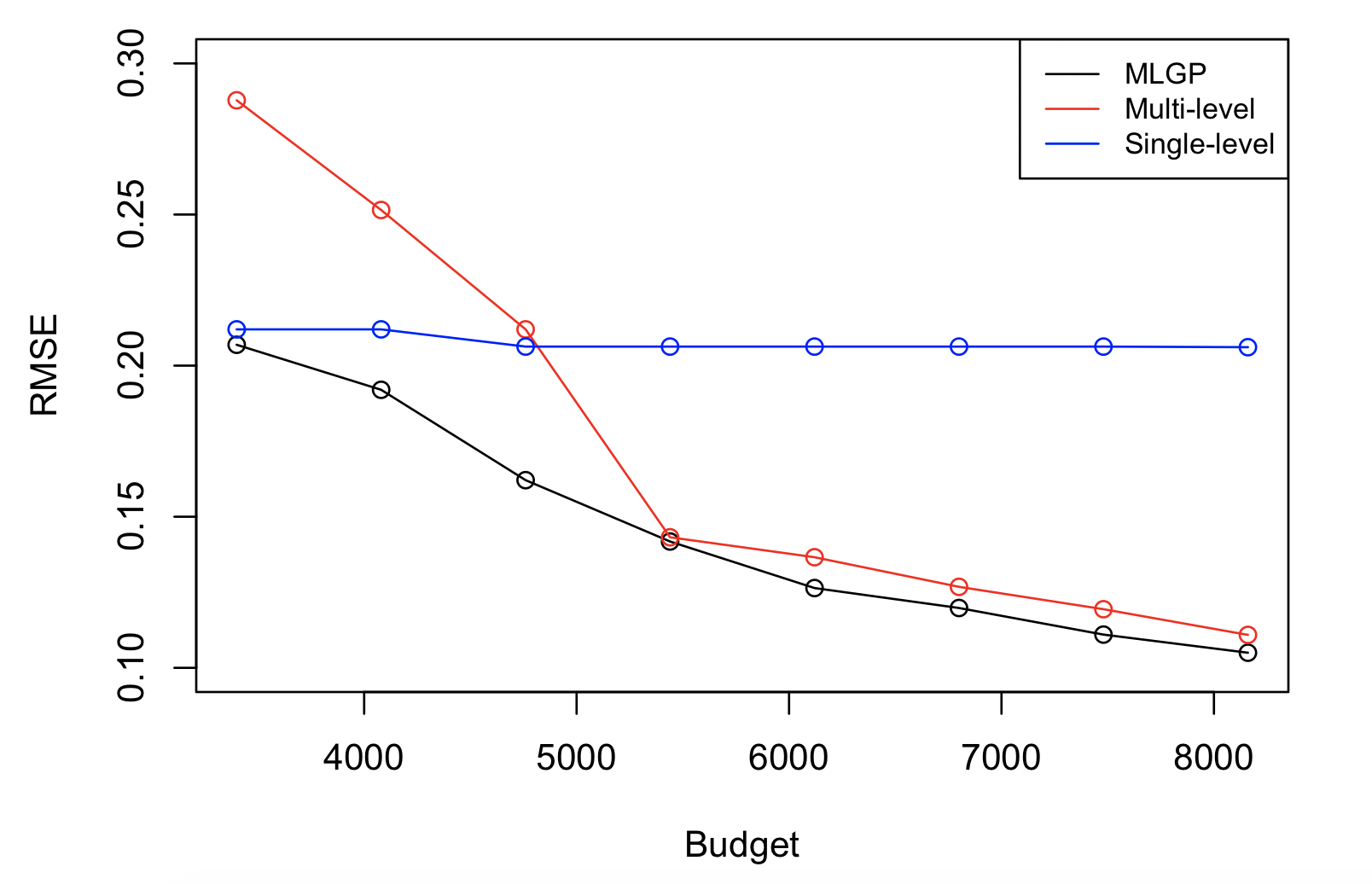}
    \caption{RMSEs of the MLGP, classical multi-fidelity method and single-level method for the 1-dimensional case under different budgets.}
    \label{fig:1dcase}
\end{figure}

Figure \ref{fig:1dcase} shows the average RMSEs of the three methods under different budgets. The blue line shows the performance of the single-level method, and it almost keeps the same even with different budgets. This is because in all cases, the best accuracy of the single-level method is achieved by only sampling at the $2$-th layer. The red line and the black line represent the performance of the multi-level method and the MLGP, respectively. One can easily see that the designs given by the MLGP lead to the emulations with the lowest RMSEs compared with other methods, which proves the superiority of the proposed MLGP.

\subsection{Numberical study for the situations with misspecified hyper-parameters}\label{mis}

We may not get the exact values for all the hyper-parameters in some cases. To show the robustness of the MLGP, we also test the performance of the proposed MLGP when we incorporate a misspecified hyper-parameter. First, we use the same parameters as the first experiment to generate the four-level function $f(x)$, i.e., $K=3$, $\lambda^2=1/3$, Mat\'ern correlation function with $\nu = 1.25$. Also, we choose the cost ratio $a=8$ and $C_a=C_b=1$. We generate designs based on the proposed MLGP using all the same parameters but a misspecified $\lambda^2=1/2$. Then we test the GPR emulation trained by this design and compare the accuracy of the testing set with other methods. Next, we use the hyper-parameters $K=3$, $\lambda^2=1/3$, Mat\'ern correlation function with $\nu = 1.25$ to generate the four-level function $f(x)$ and generate designs based on the proposed MLGP using all the same parameters but a misspecified $\nu=1.5$. GPR emulation is also used to compare the performance of the MLGP and the other two methods.

Table \ref{tab:table2} shows the results for these two experiments. One can see that the MLGP still gets the lowest error even though the misspecified hyper-parameters are incorporated.

    \begin{table}[ht]

        \caption{Mean RMSEs for the proposed MLGP method, classical multi-fidelity design method and single-level design method with misspecified $\lambda^2$ or $\nu$}
        \begin{center}
        \begin{tabular}{cccc}\hline
        \toprule
             Situation& Method & Average RMSE & Designs structure   \\
        \midrule
            & MLGP & 0.2831 & 88, 32, 21, 6 \\
            misspecified $\lambda^2$ & Multi-fidelity & 0.3441 & 56, 28, 14, 7 \\
             & Single-level & 0.3789 & 0, 0, 74, 0 \\
        \midrule
            & MLGP & 0.1652 & 88, 40, 20, 6 \\
            misspecified $\nu$ & Multi-fidelity & 0.2127 & 56, 28, 14, 7 \\
             & Single level 2& 0.2064 & 0, 0, 74, 0 \\
        \bottomrule
        \end{tabular}
        \end{center}
        \label{tab:table2}
        
    \end{table}

\subsection{Case study in two-dimensional space}\label{2d}

In this section, we will compare the performance of the proposed MLGP, the nested Latin hypercube design (NLHD), and the single-level design in a two-dimensional case. We prepare a three-level function $f(x)$ (i.e., $K=2$) defined in $[0,1]^2$ as the simulation function. Each level is assumed as a mutually independent Gaussian process (GP) with zero mean and Mat\'ern correlation function with $\nu=1.25$. We assume $\lambda^2=1/2$. In this experiment, we choose the cost ratio $a=4$ and the cost for each sample in the $0$-th layer to be $1$, i.e., $C_a=C_b=1$.

In this experiment, we assume all the true parameters are known, i.e., $\lambda^2 = 1/2$, the correlation function is  Mat\'ern correlation function with $\nu = 1.25$. For the proposed MLGP and single-level methods, we first calculate the design structure and then generate the corresponding (nested) samples using the Halton sequences. As for the NLHD, we use a design given in Table \ref{tab:table5}. Then we train a GPR emulation based on each design to test the prediction accuracy on a testing set with $500$ samples based on the root mean squared error (RMSE). In this experiment, $30$ different realizations of $f(x)$ are generated to test the average RMSE of the GPR emulations. Table \ref{tab:table4} shows the average RMSEs and the design structures of these three methods. We can see that the MLGP yields the lowest RMSE, which demonstrates the effectiveness of the MLGP in two-dimensional cases.

    \begin{table}[ht]

        \caption{Results for the proposed MLGP method, Nested Latin hypercube design (NLHD) and single level design method}
        \begin{center}
        \begin{tabular}{ccc}\hline
        \toprule
             Method &Average RMSE & Designs structure   \\
        \midrule
            MLGP & 0.1510 & 20, 7, 3 \\
              Multi-fidelity & 0.1735 & 32, 8, 2 \\
             Single-level& 0.1658 & 0, 0, 6 \\
        \bottomrule
        \end{tabular}
        \end{center}
        \label{tab:table4}
        
    \end{table}

    \begin{table}[ht]

        \caption{Nested Latin hypercube designs in two-dimensional space}
        \begin{center}
        \begin{tabular}{cc}\hline
        \toprule
             Level 3 & \makecell[c]{(0.166670,0.863640), (0.712120,0.318180),}    \\
        \midrule
            Level 2 & \makecell[c]{(0.166670,0.863640), (0.712120,0.318180), \\(0.590910,0.712120),  (0.984850,0.590910), \\(0.863640,0.984850), (0.045455,0.166670), \\(0.439390,0.045455), (0.287880,0.469700),}  \\
        \midrule
            Level 3 & \makecell[c]{(0.166670,0.863640), (0.712120,0.318180), \\(0.590910,0.712120),  (0.984850,0.590910), \\(0.863640,0.984850), (0.045455,0.166670), \\(0.439390,0.045455), (0.287880,0.469700), \\
            (0.954550,0.409090), (0.833330,0.681820), \\(0.196970,0.136360), (0.348480,0.924240), \\(0.681820,0.954550), (0.651520,0.530300),\\(0.136360,0.378790), (0.530300,0.439390),\\(0.893940,0.833330), (0.621210,0.075758),\\(0.318180,0.772730), (0.257580,0.287880),\\(0.772730,0.106060), (0.075758,0.742420),\\(0.378790,0.196970), (0.106060,0.560610),\\(0.560610,0.227270), (0.803030,0.500000),\\(0.469700,0.621210), (0.227270,0.651520),\\(0.924240,0.257580), (0.409090,0.348480),\\(0.500000,0.893940), (0.742420,0.803030),}  \\
        \bottomrule
        \end{tabular}
        \end{center}
        \label{tab:table5}
        
    \end{table}

\subsection{Case study in higher dimensional space}\label{2d}
In this section, we will compare the performance of the proposed MLGP and the single-level design in four-dimensional and eight-dimensional cases. We prepare a three-level function $f(x)$ (i.e., $K=2$) defined in $[0,1]^2$ as the simulation function. Each level is assumed as an i.i,d Gaussian process (GP) with zero mean and Mat\'ern correlation function with $\nu=1.25$. We assume $\lambda^2=1/2$. In this experiment, we choose the cost ratio $a=4$ and the cost for each sample in the $0$-th layer to be $1$, i.e., $C_a=C_b=1$.

In this experiment, we assume the available budget $B=144$, $\lambda^2 = 1/2$ and the correlation function is  Mat\'ern correlation function with $\nu = 1.25$. For the proposed MLGP and single-level methods, we first calculate the design structure and then generate the corresponding (nested) samples using the Halton sequences. Then we train a GPR emulation based on each design to test the prediction accuracy on a testing set with $500$ samples based on the root mean squared error (RMSE). In this experiment, $30$ different realizations of $f(x)$ are generated to test the average RMSE of the GPR emulations. Table \ref{tab:table5} shows the average RMSEs and the design structures of these two methods. We can see that the MLGP yields the lowest RMSE, which demonstrates the effectiveness of the MLGP in higher dimensional cases.

    \begin{table}[ht]

        \caption{Results for the proposed MLGP method and the single-level design method in higher dimensional cases}
        \begin{center}
        \begin{tabular}{ccccc}\hline
        \toprule
              &\multicolumn{2}{c}{$d=4$} & \multicolumn{2}{c}{$d=8$}   \\
        \midrule
              &Avg. RMSE & Structure&Avg. RMSE & Structure   \\
        \midrule
            MLGP & 0.3702 & 24, 10, 5 & 1.074 & 12, 9, 6 \\
            Single-level& 0.4621 & 0, 0, 9 & 1.181 & 0, 0, 9 \\
        \bottomrule
        \end{tabular}
        \end{center}
        \label{tab:table5}
        
    \end{table}

\section{Conclusion}\label{conclusion}
In this work, we propose new experimental designs for multi-fidelity computer experiments, called the multi-level Gaussian process (MLGP) designs, and prove that they are theoretically superior to any single-fidelity designs. In our theoretical analysis, we assume that a quasi-uniform nested design is used. It is not clear how to obtain such a nested design efficiently. \cite{chen2017flexible} proposed a numerical algorithm to search for good nested designs. In our work, we use Halton sequences, which are computationally efficient and nearly quasi-uniform. 
The proposed designs are readily applicable for a wide range of multi-fidelity computer experiments such as the heat exchanger problem described in \citep{qian2006building}.

\section*{Acknowledgement}
The authors are grateful for the AE and reviewers for very helpful comments.
This research is supported by NSF DMS-1914636, DMS-2312173, and CNS-2328395.

\bibliographystyle{elsarticle-num-names} 
\bibliography{ref}

\appendix

\section{Proofs for Theorem 1-3}
In this section, we prove Theorem 1-3.
\subsection{Proof of Theorem \ref{theorem1}}\label{thm1proof}

\begin{proof}

By definition,
\begin{align*}
    \hat{y}_K(x) &= \mathbb{E}\left[ \sum_{i=0}^{K}\delta_i(x) \Big| y_K(\mathcal{X}_K),...,y_0(\mathcal{X}_0) \right]\\
    &= \sum_{i=0}^{K}\mathbb{E}\left[ \delta_i(x) \Big| y_K(\mathcal{X}_K),...,y_0(\mathcal{X}_0) \right]
\end{align*}
Because $\delta_i$'s are mutually independent, for each $i$ we can get
\begin{eqnarray*}
    &&\mathbb{E}\left[ \delta_i(x) \Big| y_K(\mathcal{X}_K),\ldots,y_0(\mathcal{X}_0) \right] \\
    &=& 
    \mathbb{E}\left[ \delta_i(x) \Big| y_K(\mathcal{X}_K)-y_{K-1}(\mathcal{X}_k),\ldots,y_1(\mathcal{X}_1)-y_0(\mathcal{X}_0),y_0(\mathcal{X}_0) \right]\\
    &=&\mathbb{E}\left[ \delta_i(x) \Big| y_i(\mathcal{X}_i) -  y_{i-1}(\mathcal{X}_{i})\right].
\end{eqnarray*}
Then we can rewrite the $\hat{y}_K(x)$ as 
$$\hat{y}_K(x) = \sum_{i=0}^{K}\mathbb{E}\left[ \delta_i(x) \Big| y_i(\mathcal{X}_i) -  y_{i-1}(\mathcal{X}_{i}) \right].$$

\end{proof}

\subsection{Proof of Theorem \ref{theorem2}}\label{thm2proof}


\begin{proof}
By definition,
\begin{align*}
\mathbb{E} \left\|y_K - y_\infty\right\|^2_{L_2} &=\mathbb{E} \left\|\sum_{i=K+1}^\infty \delta_i\right\|^2_{L_2} \\
 &= \sum_{i=K+1}^\infty\mathbb{E}\left\| \delta_i\right\|^2_{L_2}\\
 &= \sum_{i=K+1}^\infty \mathbb{E}\left(\int_\mathfrak{X}\delta_i^2 dx\right)\\
 &=\sum_{i=K+1}^\infty \int_\mathfrak{X}\mathbb{E}(\delta_i^2) dx\\
 &= \sum_{i=K+1}^\infty C_s\lambda^{2i}\sigma^2\\
 &= C_l^2\lambda^{2K}\sigma^2,
\end{align*}
where $\mathfrak{X}$ denotes the input domain of $x$, $C_s$ is the area of $\mathfrak{X}$ and constant $C_l = \sqrt{C_s}/(1-\lambda)$.
\end{proof}

\subsection{Proof of Theorem \ref{theorem3}}\label{thm3proof}

\begin{proof}
Based on the Theorem 1, we have
\begin{align*}
    \mathbb{E}\left\|y_K - \hat{y}_K\right\|^2_{L_2} &= \mathbb{E}\left\|\sum_{i=0}^k(I_i(\delta_i)-\delta_i)\right\|^2_{L_2} \\
    &= \int_{\mathfrak{X}}\sum_{i=0}^k(I_i(\delta_i)-\delta_i )^2dx \\
    &= \int_{\mathfrak{X}}\sum_{i=0}^k \mathbb{E}((I_i(\delta_i)-\delta_i)^2)dx \\
    &\leq \sum_{i=0}^k p\sigma^2\lambda^{2i}n_i^{-\nu/d},
\end{align*}
where $n_i$ is the number of samples in the $i$-th step, $p$ is a constant only based on the correlation function, $d$ is the dimension of the input space and $\nu$ is a constant.
\end{proof}

\section{Solve the optimization problem}\label{ap:solve the obj}
Consider the optimization problem
\begin{alignat}{2}
\min_{n_i} \quad &\sum_{i=0}^K n_iC_i \label{ap:newnewobjectve}\\
\mbox{s.t.}\quad
&\sum_{i=0}^K \lambda^{2i}\sigma^2n_i^{-\nu/d} \leq \epsilon^2/2\label{ap:newnewconstraint}
\end{alignat}
We form the Lagrange function of (\ref{ap:newnewobjectve}) as 
\begin{eqnarray*}
    L(n_0,n_1,...,n_K, \eta) = \sum_{i=0}^K n_iC_aa^i + \eta(\sum_{i=0}^K p\lambda^{2i}\sigma^2n_i^{-\nu/d} - \epsilon^2/2).
\end{eqnarray*}
To find a local optimum, we need to have $\nabla L=0$, leading to $K+2$ conditions
\begin{eqnarray}\label{ap:npartial}
    \frac{\partial L}{\partial n_i}=C_aa^i - \frac{p\eta\nu}{d}\lambda^{2i}\sigma^2n_i^{-(\nu/d+1)}=0,\quad i=0,1,...,K
\end{eqnarray}
\begin{eqnarray}\label{ap:etapartial}
    \frac{\partial L}{\partial \eta}=\sum_{i=0}^Kp\lambda^{2i}\sigma^2n_i^{-\nu/d}-\epsilon^2/2=0.
\end{eqnarray}
Solving the $(K+1)$ equations in (\ref{ap:npartial}) leads to
\begin{eqnarray}\label{ap:n}
    n_i = (\frac{pC_ad}{\nu \sigma^2}\frac{a^i}{\lambda^{2i}}\eta^{-1})^{-\frac{d}{d+\nu}},\quad i=0,1,...,K.
\end{eqnarray}
Substitute (\ref{ap:n}) into (\ref{ap:etapartial}). Then we have
\begin{eqnarray}\label{ap:eta}
    \eta = \frac{C_ad}{\nu}(\sigma^2)^{\frac{d}{\nu}}(\frac{p\epsilon^2}{2S})^{-\frac{d+\nu}{\nu}},
\end{eqnarray}
where
\begin{align*}
    S = \sum_{i=0}^K(a^i)^{\frac{\nu}{d+\nu}}(\lambda^{2i})^{\frac{d}{d+\nu}}.
\end{align*}
Substitute (\ref{ap:eta}) into (\ref{ap:n}). We get
\begin{eqnarray*}
    n_i = (\frac{\epsilon^2}{2p\sigma^2S})^{-\frac{d}{\nu}}(\frac{a^i}{\lambda^{2i}})^{-\frac{d}{d+\nu}},\quad i=0,1,...,K.
\end{eqnarray*}
Since $n_i$ is an integer, we choose
\begin{eqnarray*}
    n_i^* = \lceil n_i \rceil
\end{eqnarray*}
to get an approximate solution to the original problem.

\section{Analysis of cost for the MLGP}\label{MLGPcostproof}
In this section, we make the analysis of the cost based on the $K$ and $n_i^*$ given in the above section. The total cost is given by
\begin{align*}
     C_\epsilon = \sum_{i=0}^K n_i^*C_i \leq \sum_{i=0}^K n_i^*C_aa^i.
\end{align*}
Since $\lceil n_i \rceil \leq n_i +1$, we have
\begin{eqnarray}\label{ap:cost}
    C_\epsilon &\leq& \sum_{i=0}^K Ca (n_i+1) a^i \nonumber\\
    &\eqsim& \sum_{i=0}^K (S \sigma^2 (\epsilon^2/2)^{-1})^{\frac{d}{\nu}} \sum_{i=0}^K(a^{\frac{\nu}{d+\nu}}\lambda^{\frac{2d}{d+\nu}})^i + \sum_{i=0}^K a^i \nonumber \\
    &=& (2\sigma^2 \epsilon^{-2})^{\frac{d}{v}} S^{\frac{d+\nu}{\nu}} +  \sum_{i=0}^K a^i. 
\end{eqnarray}
We denote the first part on the right-hand side of (\ref{ap:cost}) as part (I) and the second part as (II), and consider them separately. 

If $(a^i)^{\frac{\nu}{d+\nu}}(\lambda^{2i})^{\frac{d}{d+\nu}} = 1$ (i.e., $\alpha\nu = d\beta$), then
\begin{eqnarray}
    (I) &=& (2\sigma^2 \epsilon^{-2})^{\frac{d}{v}} (K+1)^{\frac{d+\nu}{\nu}} \nonumber\\
    &\eqsim& \epsilon^{-2d/\nu}(2\log_{\lambda^2}\frac{\epsilon}{2C_l\sigma})^{\frac{d+\nu}{\nu}}\nonumber\\
    &\eqsim& \epsilon^{-2d/\nu}|\ln \epsilon^2|^{\frac{d+\nu}{\nu}}\nonumber;\\
    \nonumber\\
    (II) &=& \frac{a^K(1-a^{-(K+1)})}{1-a^{-1}}\nonumber\\
    &\eqsim& a^K \eqsim (\epsilon^2)(\ln a/\ln \lambda^2)\nonumber\\
    &=& \epsilon^{-2d/\nu}.\nonumber
\end{eqnarray}

Since $\epsilon^{-2d/\nu}|\ln \epsilon^2|^{\frac{d+\nu}{\nu}} > \epsilon^{-2d/\nu}$, $C_\epsilon \lesssim \epsilon^{-2d/\nu}|\ln \epsilon^2|^{\frac{d+\nu}{\nu}}$. 

If $(a^i)^{\frac{\nu}{d+\nu}}(\lambda^{2i})^{\frac{d}{d+\nu}} < 1$ (i.e., $\alpha\nu < d\beta$), $S$ will converge to a limitation which is independent of $K$. As a result, we have
\begin{eqnarray}
    (I)\lesssim \epsilon^{-2d/\nu}\nonumber;\\
    \nonumber\\
    (II)\lesssim \epsilon^{-2\alpha/\beta}\nonumber.
\end{eqnarray}
Since $d/\nu > \alpha/\beta$, $C_\epsilon \lesssim \epsilon^{-2d/\nu}$. 

If $(a^i)^{\frac{\nu}{d+\nu}}(\lambda^{2i})^{\frac{d}{d+\nu}} > 1$ (i.e., $\alpha\nu > d\beta$), we have
\begin{eqnarray}
    (I) &\lesssim& \epsilon^{-2d/\nu} (\exp{\frac{\alpha\nu - d\beta}{d+\nu}})^{\frac{K(d+\nu)}{\nu}}\nonumber\\
    &\lesssim& \epsilon ^{-2d/\nu} \exp{(-\frac{(\ln \epsilon^2 (\alpha\nu - d\beta))}{\nu\beta})}\nonumber\\
    &=&\epsilon^{-2\alpha/\beta}\nonumber;\\
    \nonumber\\
    (II) &\eqsim& \epsilon^{-2\alpha/\beta}\nonumber.
\end{eqnarray}
As a result, $C_\epsilon \lesssim \epsilon^{-2\alpha/\beta}$. 

In summary, 
\begin{eqnarray*}
    C_\epsilon\lesssim\left\{\begin{matrix}
 \epsilon^{-2d/\nu},&\alpha\nu < d\beta \\ 
 \epsilon^{-2d/\nu}|\ln \epsilon^2|^{\frac{d+\nu}{\nu}}, & \alpha\nu = d\beta\\ 
 \epsilon^{-2\alpha/\beta},&\alpha\nu > d\beta  
\end{matrix}\right.
\end{eqnarray*}

\section{Analysis of cost for single-level}\label{SLcostproof}
The error caused by single-level approximation can be bounded as
\begin{eqnarray*}
    \mathbb{E}(\left\|y_\infty - \hat{y}_K^{SL}\right\|^2) = \mathbb{E}(\left\|y_\infty - y_K^{SL}\right\|^2 + \left\|y_K^{SL} - \hat{y}_K^{SL}\right\|^2),
\end{eqnarray*}
where $\hat{y}_K^{SL} = I(\delta_K)$. Then the objective function for this single-level approximation can be summarized as
\begin{align}{2}
\min_{K, n_K}  \quad & n_KC_K \label{ap:objectvesl}\\
\mbox{s.t.}\quad
&\mathbb{E}\left\|y_\infty - \hat{y}_K^{SL}\right\|^2_{L^2} \leq \epsilon\label{ap:constraint1}
\end{align}

Based on Theorems 1 and 2, we have 
\begin{eqnarray*}\label{ap:costsl}
    \mathbb{E}\left\|y_\infty - \hat{y}_K^{SL}\right\|^2 \gtrsim \lambda^{2K}\sigma^2 + \sigma^2n_K^{-\nu/d},
\end{eqnarray*}
where $n_K$ is the number of samples. It should be sufficient to bound the upper bound of the error by $\epsilon$, that is, 
\begin{eqnarray}\label{ap:solvefornk}
    \lambda^{2K} + n_K^{-\nu/d} \lesssim \epsilon^2.
\end{eqnarray}

From the (\ref{ap:solvefornk}), we can get the lower bound for $n_K$, that is,
\begin{eqnarray*}
    n_K \gtrsim (\epsilon^2 - \lambda^{2K})^{-d/\nu}.
\end{eqnarray*}
Then we have the lower bound for the cost of single-level approximation:
\begin{eqnarray}\label{ap:newcostsl}
    C_\epsilon^{SL} = n_KC_k \gtrsim a^K (\epsilon^2 - \lambda^{2K})^{-d/\nu}.
\end{eqnarray}
We denote $a = \exp{(\alpha)}$ and $\lambda^2 = \exp{(-\beta)}$ for convenience, $\alpha, \beta>0$. Then wen can rewrite (\ref{ap:newcostsl}) as
\begin{eqnarray}\label{lnnewcostsl}
    \ln C_\epsilon^{SL} \gtrsim \alpha K - \frac{d}{\nu}\ln {(\epsilon^2 - e^{-\beta K})}.
\end{eqnarray}
Now we want to find the lower bound of $C_\epsilon^{SK}$, that is, we need to find the lower bound for the right-hand side of (\ref{lnnewcostsl}). Consider the following function with respect to $K$: 
\begin{eqnarray*}
    T(K) = \alpha K - \frac{d}{\nu}\ln {(\epsilon^2 - e^{-\beta K})}.
\end{eqnarray*}
Set the derivative of $T$ with respect to $K$ equal to $0$:
\begin{eqnarray*}
    \frac{dT}{dK} &=& \alpha  + \frac{d\beta}{\nu} (1 - \frac{\epsilon^2}{\epsilon^2 - e^{-\beta K}})=0.
\end{eqnarray*}
We can get the solution $K_0 =-\frac{1}{\beta} \ln \frac{\epsilon^2 \alpha \nu}{\alpha\nu + \beta d}$. When $K<K_0$, $T(K)$ is decreasing and when $K>K_0$, $T(K)$ is increasing. Substitute $K=0$ into $T(K)$, we can get the lower bound of the cost for the single-level:
\begin{eqnarray*}
    C_\epsilon^{SL} \gtrsim \epsilon^{-2(\frac{d}{\nu} + \frac{\alpha}{\beta}) }.
\end{eqnarray*}

\end{document}